\documentclass[12pt]{iopart}

\usepackage{iopams}  
\usepackage[final]{epsfig}
\usepackage{bm}
\usepackage{setspace}

\newcommand{\Tc}{T_\mathrm{c}}

\newcommand{\ecc}{\varepsilon}

\newcommand{\dNdy}{dN_\mathrm{ch}/dy}

\newcommand{\be}[1]{\begin{equation}\label{#1}}
\newcommand{\ee}{\end{equation}}

\newcommand{\eq}{{\,=\,}}

\def\La{\langle}
\def\Ra{\rangle}

\begin{document}

\title[]{The QGP shear viscosity -- elusive goal or just around the
corner?}
\author{\underline{Chun Shen}$^1$, Steffen A Bass$^2$, Tetsufumi Hirano$^{3,4}$,
        Pasi Huovinen$^5$, Zhi Qiu$^1$, Huichao Song$^6$ and Ulrich Heinz$^1$} 
\address{$^1$\,Department of Physics, The Ohio State University,
             Columbus, Ohio 43026, USA\\ 
         $^2$\,Department of Physics, Duke University,
             Durham, North Carolina 27708, USA\\ 
         $^3$\,Department of Physics, Sophia University, 
             Tokyo 102-8554, Japan\\                      
         $^4$\,Department of Physics, The University of Tokyo, 
             Tokyo 113-0033, Japan\\ 
         $^5$\,ITP, J.W.Goethe-Universit\"at,
               D-60438 Frankfurt a.\,M., Germany\\
         $^6$\,Lawrence Berkeley National Laboratory, 
               Berkeley, California 94720, USA              
}
%
%
\begin{abstract}
With the new viscous hydrodynamic + hadron cascade hybrid code {\tt VISHNU},
a rather precise (${\cal O}(25\%)$) extraction of the QGP shear viscosity 
$(\eta/s)_\mathrm{QGP}$ from heavy-ion elliptic flow data is possible 
{\em if} the initial eccentricity of the collision fireball is 
known with $<5\%$ accuracy. At this point, eccentricities from initial state
models differ by up to 20\%, leading to an ${\cal O}(100\%)$ uncertainty
for $(\eta/s)_\mathrm{QGP}$. It is shown that a simultaneous comparison 
of elliptic and triangular flow, $v_2$ and $v_3$, puts strong constraints
on initial state models and can largely eliminate the present uncertainty
in $(\eta/s)_\mathrm{QGP}$. The variation of the differential elliptic
flow $v_2(p_T)$ for identified hadrons between RHIC and LHC energies provides 
additional tests of the evolution model.   
\end{abstract}
%
\vspace*{-3mm}

\noindent
{\bf Prologue -- how to measure $\bm{(\eta/s)_\mathrm{QGP}}$:}
Hydrodynamics converts the initial spatial deformation of the fireball 
created in relativistic heavy-ion collisions into final state momentum 
anisotropies. Viscosity degrades the conversion efficiency
$\ecc_x\eq\frac{\La\!\La y^2{-}x^2\Ra\!\Ra}{\La\!\La y^2{+}x^2\Ra\!\Ra}
\to 
\ecc_p\eq\frac{\La T^{xx}{-}T^{yy}\Ra}{\La T^{xx}{+}T^{yy}\Ra}$ of the
fluid; for given initial fireball ellipticity $\ecc_x$, the viscous 
suppression of the dynamically generated total momentum anisotropy 
$\ecc_p$ is monotonically related to the specific shear viscosity $\eta/s$. 
The observable most directly related to $\ecc_p$ is the total charged 
hadron elliptic flow $v_2^\mathrm{ch}$ \cite{Heinz:2005zg}. Its 
distribution in $p_T$ depends on the chemical composition and 
$p_T$-spectra of the various hadron species; the latter evolve in
the hadronic stage due to continuously increasing radial 
flow (and so does $v_2(p_T)$), even if (as expected at top LHC energy 
\cite{Hirano:2007xd}) $\ecc_p$ fully saturates in the QGP phase. When 
(as happens at RHIC energies) $\ecc_p$ does not reach saturation before 
hadronization, dissipative hadronic dynamics \cite{Hirano:2005xf} affects 
not only the distribution of $\ecc_p$ over hadron species and $p_T$, but 
even the final value of $\ecc_p$ itself, and thus of $v_2^\mathrm{ch}$ from 
which we want to extract $\eta/s$. To isolate the QGP viscosity
$(\eta/s)_\mathrm{QGP}$ we therefore need a hybrid code that couples 
viscous hydrodynamics of the QGP to a realistic model of the late hadronic 
stage, such as {\tt UrQMD} \cite{Bass:1998ca}, that describes its dynamics 
microscopically. {\tt VISHNU} \cite{Song:2010aq} is such a code.

\begin{figure}[htb]
\begin{center}
 \includegraphics[width=0.35\linewidth,clip=,angle=0]{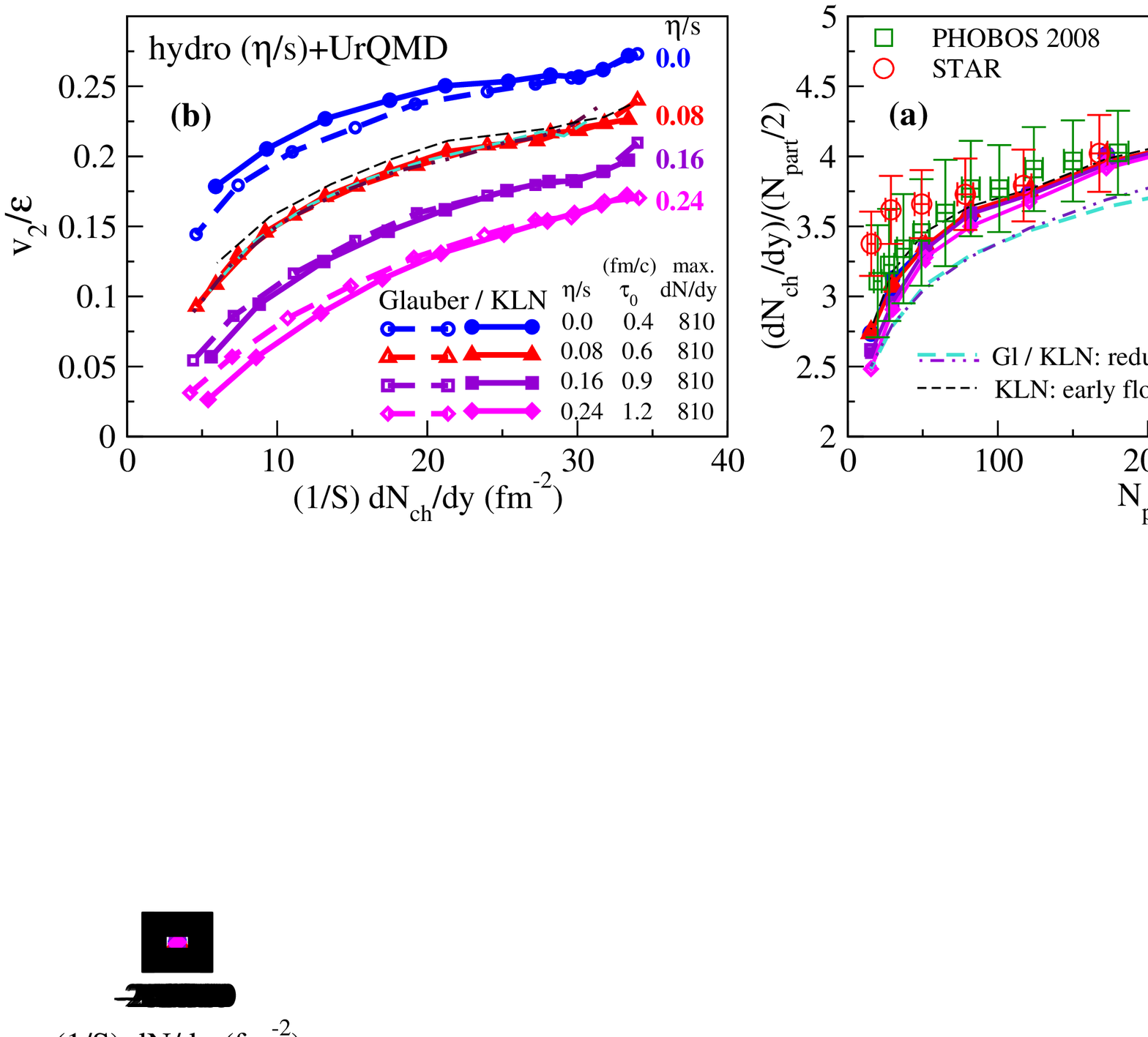}  
 \includegraphics[width=0.635\linewidth,clip=,angle=0]{Fig2.eps}  
\caption{\label{F1} (Color online) Centrality dependence of 
eccentricity-scaled elliptic flow \cite{Song:2010mg}.
\vspace*{-8mm}        
}
\end{center}
\end{figure}

\bigskip
\noindent
{\bf Extraction of $\bm{(\eta/s)_\mathrm{QGP}}$ from 200\,$\bm{A}$\,GeV
Au+Au collisions at RHIC:}
The left panel in Fig.~\ref{F1} shows that such an approach yields a
universal dependence of the ellipticity-scaled total charged hadron 
elliptic flow, $v_2^\mathrm{ch}/\ecc_x$, on the charged hadron multiplicity 
density per overlap area, $(1/S)(\dNdy)$, that depends only on 
$(\eta/s)_\mathrm{QGP}$ but not on the details of the initial state
model that provides $\ecc_x$ and $S$ \cite{Song:2010mg}. Pre-equilibrium 
flow and bulk viscous effects on these curves are small \cite{Song:2010mg}. 

The QGP viscosity can be extracted from experimental $v_2^\mathrm{ch}$ data 
by comparing them with these universal curves. The right panels of 
Fig.~\ref{F1} show this for MC-Glauber and MC-KLN initial state models 
\cite{Song:2010mg}. In both cases the slope of the data 
\cite{Ollitrault:2009ie} is correctly reproduced (not true for ideal 
nor viscous hydrodynamics with constant $\eta/s$). Due to the ${\sim}20\%$ 
larger ellipticity of the MC-KLN fireballs, the magnitude of 
$v_\mathrm{2,exp}^\mathrm{ch}/\ecc_x$ differs between the two models. 
Consequently, the value of $(\eta/s)_\mathrm{QGP}$ extracted from this 
comparison changes by more than a factor 2 between them. Relative to the 
initial fireball ellipticity all other model uncertainties are negligible. 
Without constraining $\ecc_x$ more precisely, 
$(\eta/s)_\mathrm{QGP}$ cannot be determined to better than a factor 2 from 
elliptic flow data alone, irrespective of any other model improvements. 
Taking the MC-Glauber and MC-KLN models to represent a reasonable range of 
initial ellipticities, Fig.~\ref{F1} gives 
$1{\,<\,}4\pi(\eta/s)_\mathrm{QGP}{\,<\,}2.5$ for 
temperatures $\Tc{\,<\,}T{\,<\,}2\Tc$ probed at RHIC. 

%
\begin{figure}[htb]
\begin{center}
\hspace*{1cm}
 \includegraphics[width=0.4\linewidth,clip=,angle=0]%
                 {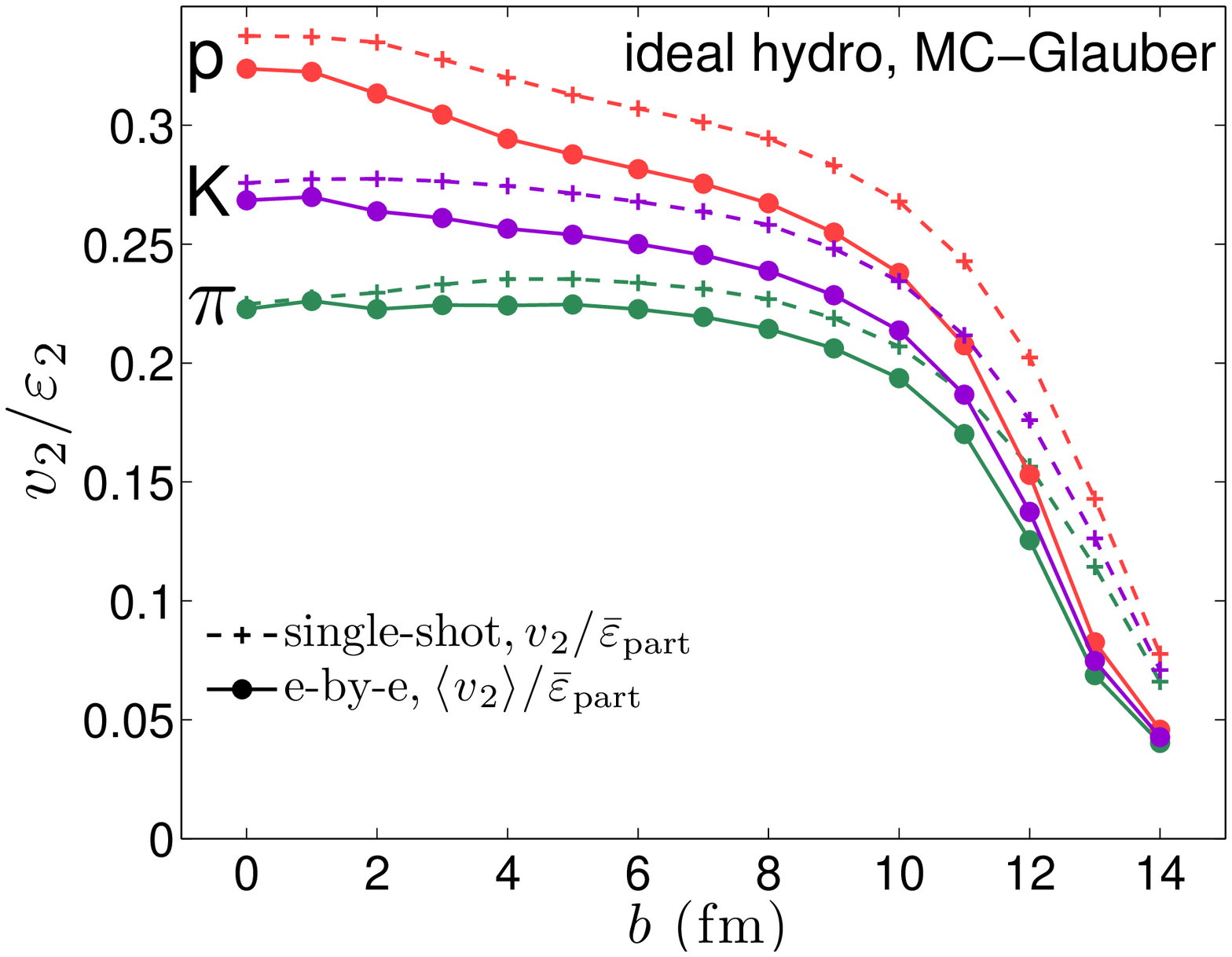}  
 \includegraphics[width=0.4\linewidth,clip=,angle=0]%
                 {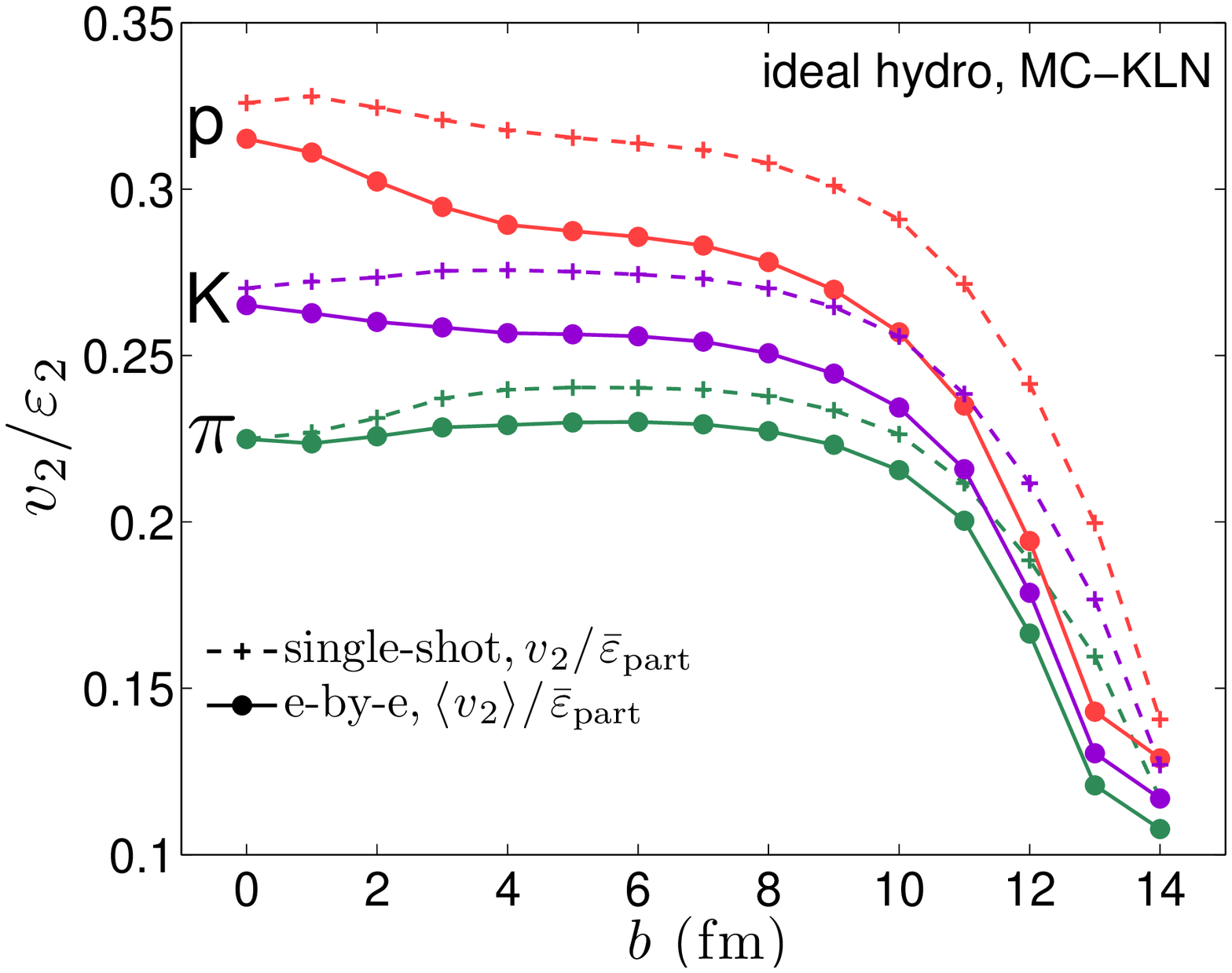}  
\caption{\label{F2} (Color online) Eccentricity-scaled elliptic flow
as function of impact parameter for pions, kaons and protons from 
single-shot and event-by-event ideal fluid evolution of fluctuating
initial conditions from the MC-Glauber (left) and MC-KLN (right) models.
\vspace*{-8mm}        
        }
\end{center}
\end{figure}
%
{\tt VISHNU} with $(\eta/s)_\mathrm{QGP}\eq\frac{1} {4\pi}$ for MC-Glauber 
and $\frac{2}{4\pi}$ for MC-KLN provides an excellent description of all 
aspects of soft ($p_T{\,<\,}1.5$\,GeV) hadron production ($p_T$-spectra 
and differential $v_2(p_T)$ for all charged hadrons together as well as 
for individual identified species) in 200\,$A$\,GeV Au+Au collisions at 
all but the most peripheral collision centralities \cite{Song:2011hk}. 
Such a level of theoretical control is unprecedented.  

\bigskip
\noindent
{\bf Event-by-event hydrodynamics of fluctuating fireballs:} 
In Fig.~\ref{F1} we evolved a smooth averaged initial profile 
(``single-shot hydrodynamics''). This overestimates the conversion 
efficiency $v_2/\ecc$ \cite{Andrade:2006yh,Qiu:2011iv}. Fig.~\ref{F2} 
shows that event-by-event ideal fluid dynamical evolution of fluctuating 
fireballs reduces $v_2/\ecc$ by a few percent \cite{Qiu:2011iv}. The 
effect is only 
$\sim5\%$ for pions but larger for heavier hadrons. We expect it to be less 
in viscous hydrodynamics which dynamically dampens large initial fluctuations. 
A reduced conversion efficiency $v_2/\ecc$ from event-by-event evolution will 
reduce the value of $(\eta/s)_\mathrm{QGP}$ extracted from $v_2^\mathrm{ch}$; 
based on what we see in ideal fluid dynamics, the downward shift for 
$(\eta/s)_\mathrm{QGP}$ will at most be of order 0.02-0.03.

\begin{figure}[h!]
\begin{center}
  \includegraphics[width=0.39\linewidth,height=5cm,clip=,angle=0]%
                  {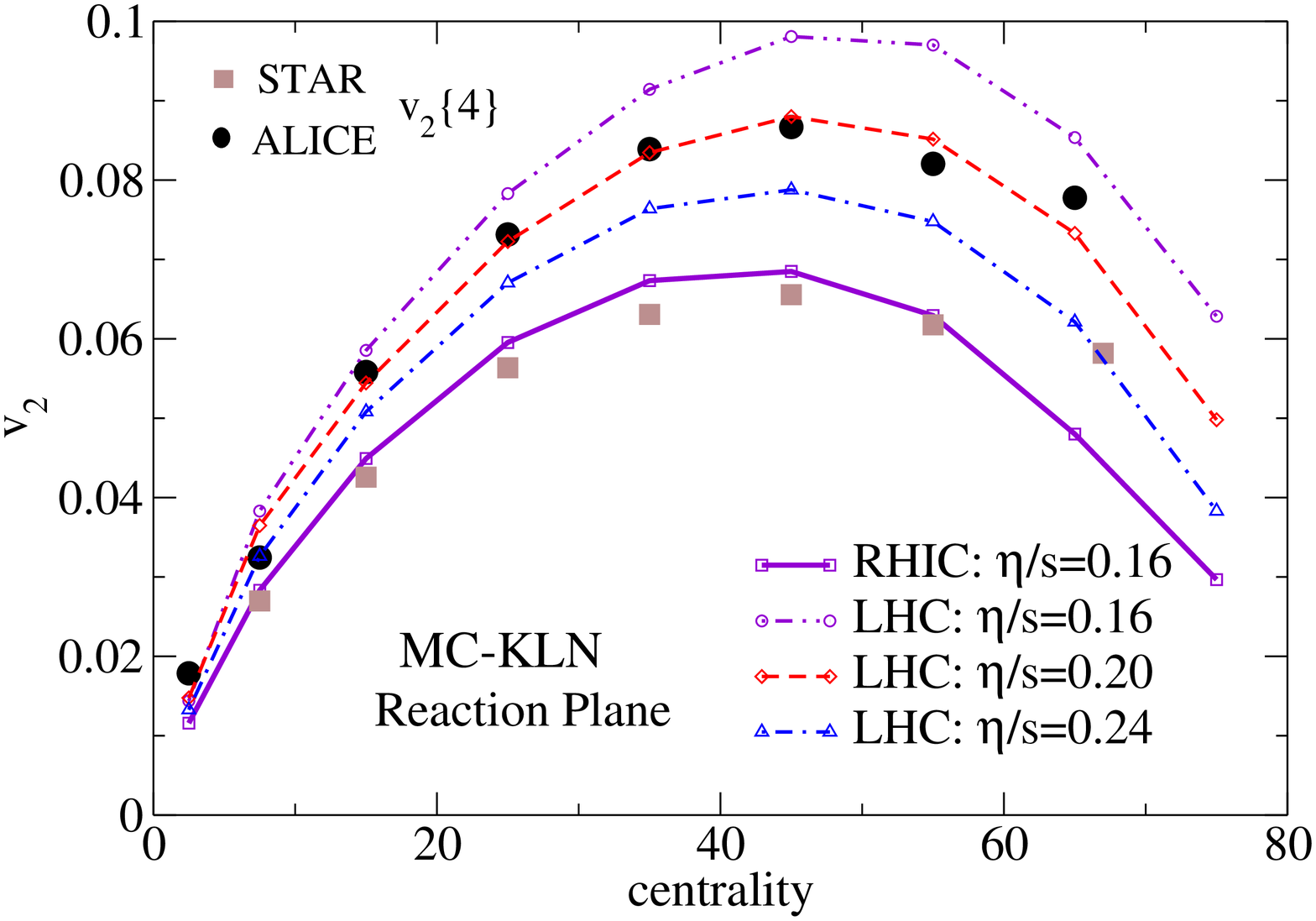}
  \includegraphics[width=0.41\linewidth,clip=,angle=0]{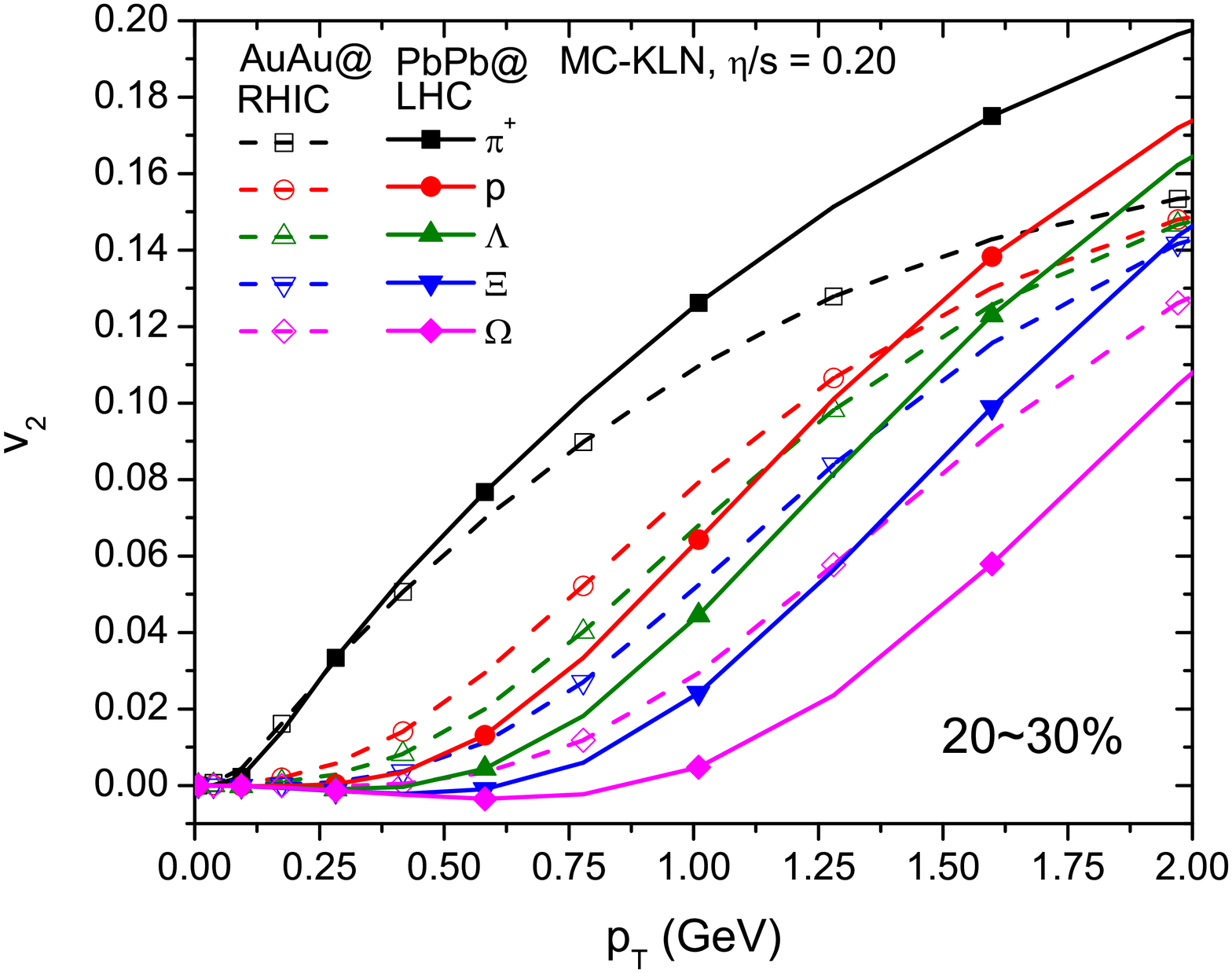}  
\caption{\label{F3} (Color online) Total charged hadron elliptic flow as
function of centrality ({\tt VISHNU}, left \cite{Song:2011qa}) and 
differential elliptic flow for identified hadrons for 20-30\% centrality 
({\tt VISH2+1}, right \cite{Shen:2011eg}) for 200\,$A$\,GeV Au+Au collisions 
at RHIC and 2.76\,$A$\,TeV Pb+Pb collisions at the LHC. Experimental data 
are from \cite{Aamodt:2010pa}.
\vspace*{-5mm}        
        }
\end{center}
\end{figure}

\medskip
\noindent
{\bf Predictions for spectra and flow at the LHC:} 
The successful comprehensive fit of spectra and elliptic flow at RHIC 
\cite{Song:2011hk} allows for tightly constrained LHC predictions. 
Fig.~\ref{F3} shows such predictions for both pure viscous hydrodynamics 
{\tt VISH2+1} \cite{Shen:2011eg} and {\tt VISHNU} \cite{Song:2011qa}. 
%
%
A straightforward extrapolation with fixed $(\eta/s)_\mathrm{QGP}$ 
overpredicts the LHC $v_2^\mathrm{ch}$ values by 10-15\%; a slight increase
of $(\eta/s)_\mathrm{QGP}$ from 0.16 to 0.20 (for MC-KLN) gives better 
agreement with the ALICE data \cite{Aamodt:2010pa}. However, at LHC 
energies $v_2$ becomes sensitive to details of the initial shear stress 
profile \cite{Shen:2011eg}, and no firm conclusion can be drawn yet 
whether the QGP turns more viscous (i.e. less strongly coupled) at 
higher temperatures. The right panel shows that, at fixed 
$p_T{\,<\,}1$\,GeV, $v_2(p_T)$ increases from RHIC to LHC for pions but 
decreases for all heavier hadrons. The similarity at RHIC and LHC of 
$v_2^\mathrm{ch}(p_T)$ for the sum of all charged hadrons thus appears 
accidental.

\medskip
\noindent
{\bf Constraining initial state models by simultaneous measurement of 
$\bm{v_2}$ and $\bm{v_3}$:}  
While the ellipticities $\ecc_2$ differ by about 20\% between MC-KLN and 
MC-Glauber models, their triangularities
%
\begin{figure}[htb]
\begin{center}
  \includegraphics[width=0.40\linewidth,clip=,angle=0]{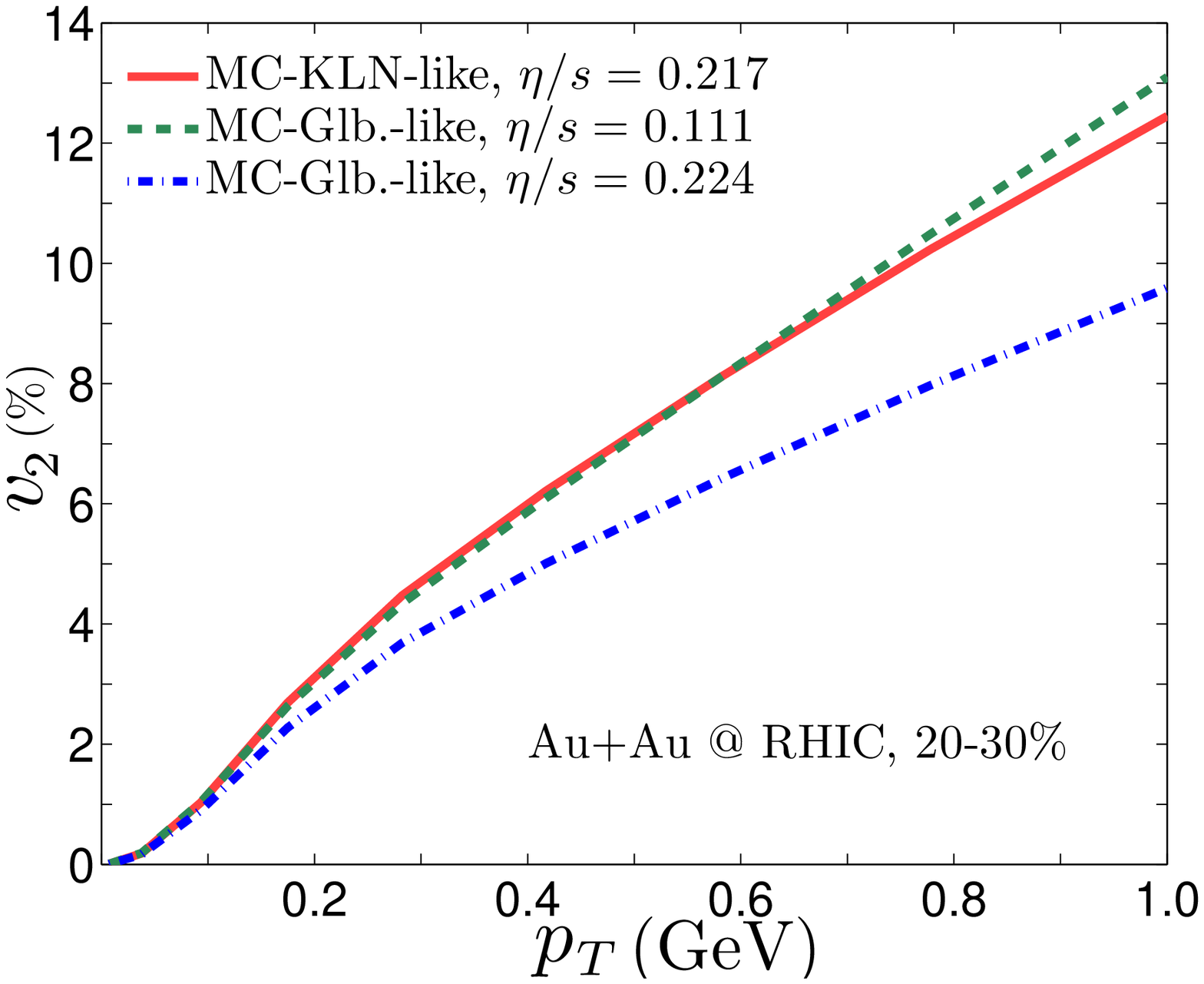}
  \includegraphics[width=0.39\linewidth,clip=,angle=0]{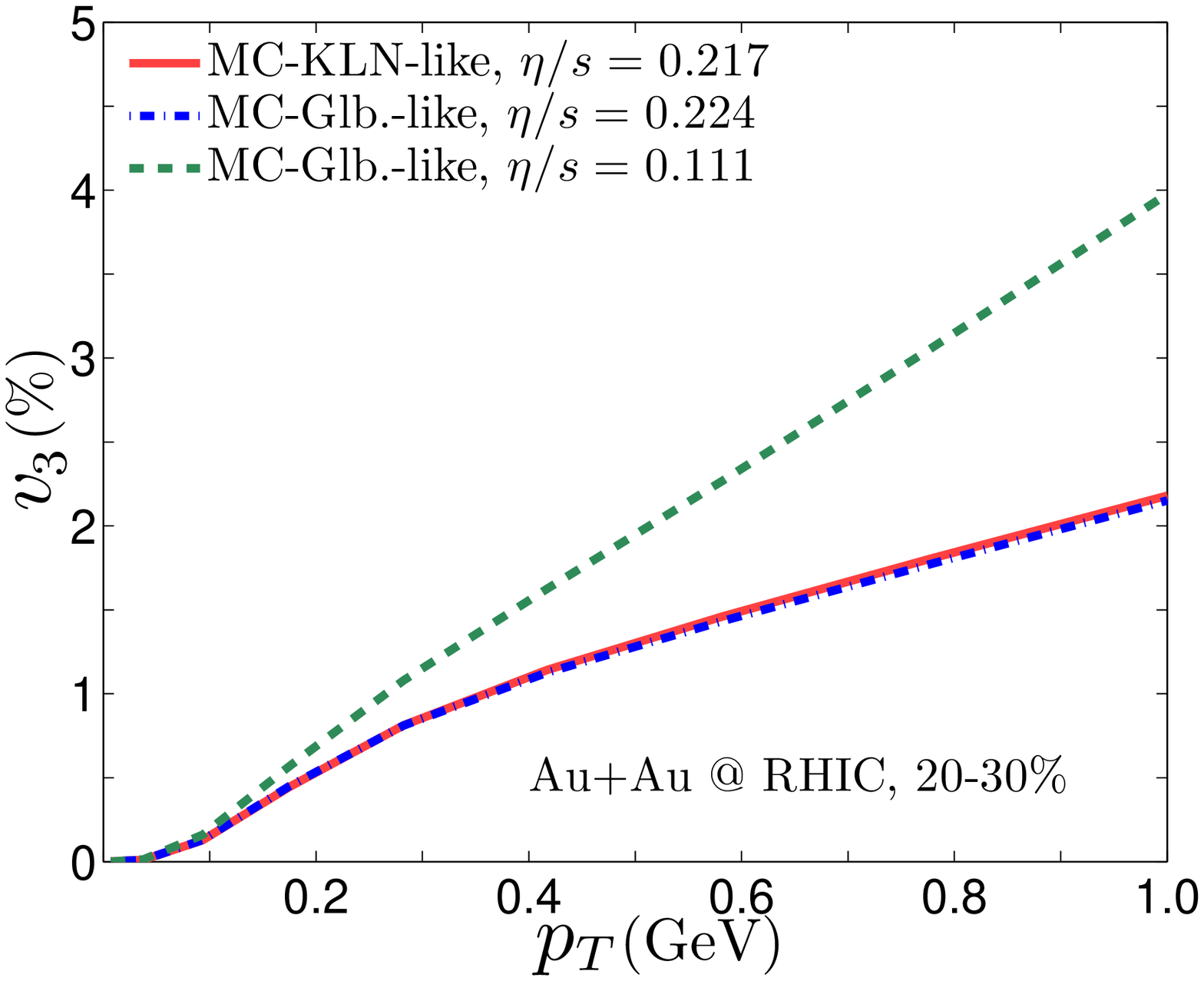} 
\caption{\label{F4} (Color online) $p_T$-differential elliptic and 
triangular flow from viscous hydrodynamics for initial eccentricities 
from the MC-KLN and MC-Glauber models. 
\vspace*{-5mm}        
        }
\end{center}
\end{figure}
%
$\ecc_3$ (which are entirely due to event-by-event fluctuations) are almost 
identical \cite{Qiu:2011iv}. This suggests to use triangular flow $v_3$ 
(which is almost entirely \cite{Qiu:2011iv} driven by $\ecc_3$) to obtain a 
model-independent measurement of $(\eta/s)_\mathrm{QGP}$. Fig.~\ref{F4} 
shows $v_2^\pi(p_T)$ and $v_3^\pi(p_T)$ for deformed Gaussian fireballs with 
average eccentricities $\ecc_2$ and $\ecc_3$ (with random relative angle) 
taken from the fluctuating Glauber (``MC-Glauber-like'') and 
KLN (``MC-KLN-like'') models. It demonstrates that a given set of
flow data requires shear viscosities that differ by a factor 2 to reproduce
$v_2(p_T)$ and but the same shear viscosities in both models 
to reproduce $v_3(p_T)$. A good fit by both models to $v_2(p_T)$ produces 
dramatically different curves for $v_3(p_T)$, and {\it vice versa}. The figure 
illustrates the strong discriminating power for such simultaneous studies 
and gives hope for a much more precise extraction of $(\eta/s)_\mathrm{QGP}$ 
in the near future.    
  
\bigskip
\begin{spacing}{0.8}
{\footnotesize
\noindent
{\bf Acknowledgments:} 
This work was supported by the U.S.\ Department of Energy under grants 
No. DE-AC02-05CH11231, DE-FG02-05ER41367, \rm{DE-SC0004286}, and (within 
the framework of the JET Collaboration) \rm{DE-SC0004104}; by the Japan 
Society for the Promotion of Science through Grant-in-Aid for Scientific 
Research No. 22740151; by the ExtreMe Matter Institute (EMMI); and by 
BMBF under project No. 06FY9092. We gratefully acknowledge extensive 
computing resources provided to us by the Ohio Supercomputer Center. 
C.~Shen thanks the {\it Quark Matter 2011} organizers for support. 
}
\end{spacing}

\end{document}